\documentclass[slac_one]{revtex4}
\usepackage{graphicx}
\usepackage{fancyhdr}
\begin{document}

\title{{\small{2005 ALCPG \& ILC Workshops - Snowmass,
U.S.A.}}\\ 
\vspace{12pt}
CPT violation in the top sector} 

%

\author{J. A. R. Cembranos}
\affiliation{Department of Physics and Astronomy,
 University of California, Irvine, CA 92697 USA}
\author{A. Rajaraman}
\affiliation{Department of Physics and Astronomy,
 University of California, Irvine, CA 92697 USA}
\author{F. Takayama}
\affiliation{Institute for High-Energy Phenomenology,
Cornell University, Ithaca, NY 14853, USA}
 
\begin{abstract}
We study the viability of observation of CPT violation in the top sector at future colliders. We show possible studies and different estimates for hadronic and linear colliders. In particular, we will present current constraints for Tevatron and prospects for the LHC and the ILC. 

\end{abstract}

\maketitle

\thispagestyle{fancy}


\section{INTRODUCTION}

 Symmetries are fundamental to our physical understanding of our Universe. In addition to continuous symmetries, there are discrete symmetries such as the charge conjugation symmetry C,  parity  P,  and the time reversal symmetry T, which reverses the direction of time. We can also build other symmetries by taking products. For instance, CP is the product of the charge conjugation and parity and CPT is the product of CP and T.  

For a long time, physical laws were thought to conserve C, P and T , but experiments have shown us that the Universe is not so symmetric. Indeed, C and P are maximally violated in weak interactions  \cite{P,PDG} and evidences of CP non conservation (in 1964 \cite{CP,PDG}) and T violation (much more recently \cite{T,PDG}) have been observed in the neutral kaon system. On the other hand, we have not observed any signature of CPT non conservation so far. In fact, our present understanding of the interaction of the Universe can be summarized in Table \ref{forces}. The electroweak sector is the only one which has a non trivial behavior in relation to discrete symmetries, and precisely this sector is the last piece of the Standard Model that needs to be confirmed. This is one of the main analyses which are going to be performed by the LHC and the ILC.

\begin{table}[ht]
\centering
\begin{tabular}{||c|| c | c | c | c ||}
\hline\hline
Interactions & Gravitational & Electromagnetic & Strong & Electroweak    \cr
\hline\hline
Relative Magnitude & $10^0$  & $10^{38}$              & $10^{40}$     & $10^{15}$      \cr
\hline
Range        & $\infty$      & $\infty$               & $10^{-15}$ m  & $10^{-18}$ m   \cr
\hline
$P$-conservation    & YES & YES & YES  & NO \cr      
\hline
$T$-conservation    & YES & YES & YES  & NO \cr   
\hline
$C$-conservation    & YES & YES & YES  & NO \cr   
\hline
$CP$-conservation   & YES & YES & YES  & NO \cr     
\hline
$CPT$-conservation  & YES & YES & YES  & YES \cr
\hline\hline
\end{tabular}
\caption{\label{forces} Main characteristics of the interactions which govern the
behaviour of particles in the Standard Model.}
\end{table}

 CPT symmetry is guaranteed by the CPT theorem based on three very fundamental assumptions: any local theory, which is invariant under  Lorentz transformations and defined by a Hermitian Hamiltonian conserves CPT \cite{local}. However, many well motivated theories can produce CPT violation \cite{CPTmodels}. String theory, which is our best theory of quantum gravity, is intrinsically nonlocal and can violate CPT. Recent theories of ghost condensation can also produce CPT violation.
 Thus it may be very interesting to look for CPT violation in future collider experiments. 
 
CPT conservation implies that masses and lifetimes of particles and anti-particles are the same. Any mass difference between a particle and its antiparticle is unambiguous evidence of CPT violation. Here we  will focus on measurements of the quantity $R_{CPT}(t)\equiv 2(m_t-m_{\bar t})/(m_t+m_{\bar t})$ which parametrizes CPT violation in the top sector. 

\section{HADRONIC COLLIDERS}

We now analyze the top anti-top production at hadronic colliders. We start with the study of the di-lepton channel, where the W bosons decay leptonically. We can reconstruct the top or anti-top mass by using the invariant mass associated to the lepton and b quark coming from the  decay of the top or anti-top. The mass distribution from data coming from top and anti-top decays should have two different peaks if the CPT violation is large enough.

\begin{figure}[ht]
 \centerline{ 
    \includegraphics[width=0.48\textwidth,clip]{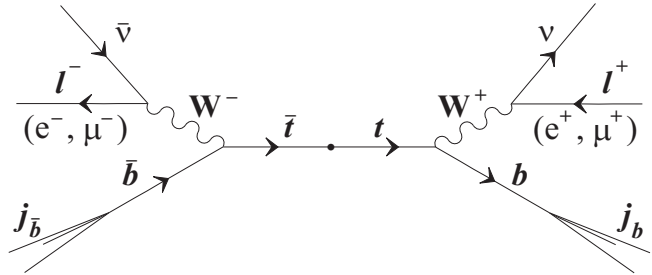}} 
  \caption{Schematics of the top and anti-top decays in the dilepton channel.} 
  \label{dileptonSch} 
\end{figure} 

We can estimate the present constraints on $R_{CPT}$  by using the Tevatron data accumulated at Fermilab from 1992 through 1995 \cite{CDFdilept1}. The analysis performed by CDF by using this technique is consistent with only one peak at: $m_t=m_{\bar t}=163\pm 2 (stat.)\pm 9 (syst.)$ GeV. We can estimate the constraint $|R_{CPT}(t)|<0.13$ at the 95\% c.l. \cite{CRT} by taking a conservative approach where we add the systematic uncertainties. 

We can also evaluate  the sensitivity of the LHC. The expected statistical uncertainty in the top mass has been estimated as 0.9 GeV, whereas the systematic errors are expected to be around 2 GeV \cite{mtdetLHC}. The LHC could thus be sensitive to $R_{CPT}(t)=0.03$ at the 95\% c.l. using this channel (supposing $\bar m_{t\bar t}\equiv (m_t-m_{\bar t})/2= 174.3$ \cite{PDG}).

A more promising signal is provided by the lepton plus jet channel, in which one of the W bosons decay leptonically whereas the other one decays hadronically. We can perform the analogous analysis to that of the di-lepton channel, reconstructing the masses with the invariant mass $m_{jjb}$ associated to the hadronic decay. 

Combining the CDF \cite{CDFl+jets} and DO data \cite{D0l+jets} gives a stronger bound on $R_{CPT}(t) < 9.2\times 10^{-2}$. The sensitivity of the LHC is also better in this channel since both statistical and systematic uncertainties are expected to be improved. Indeed, the LHC will be able to test the CPT violation of the top quark at almost one order of magnitude better than the present constraints: $|R_{CPT}(t)|\simeq 0.014$ at the 95\% c.l. or equivalently, $m_t-m_{\bar t}\simeq 2.4$ GeV \cite{CRT}. 

\section{LINEAR COLLIDERS}

The same analyses can be performed with the International Linear Collider. There are fewer studies about the determination of the top mass through top anti-top quark production, but  the statistical uncertainties will increase while the  systematic errors can be reduced \cite{Biernacik:2003xv}.  The systematic ones dominate (at least in our conservative approach) and this leads to a small improvement of the sensitivity as compared to the LHC. 

\begin{figure}[ht]
 \centerline{ 
    \includegraphics[width=0.53\textwidth,clip]{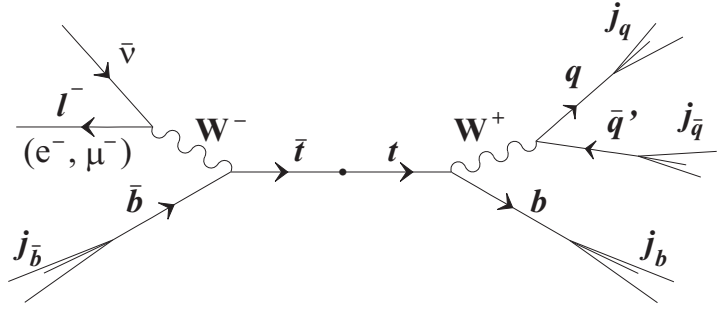}} 
  \caption{Schematic example of the top and anti-top decays in the lepton plus jets channel.} 
  \label{fig:avto1} 
\end{figure} 
\section{CONCLUSIONS}

There are other possible analyses that can be performed. It seems very interesting to study the threshold scan analysis for the production of top anti-top production in linear colliders since it is extremely sensitive to the top quark mass. However this method is not so promising in relation to the observation of CPT violation. Other techniques to reconstruct the top mass could also be very interesting, such as the analysis of the J/$\psi$ from the $b$ decay in the LHC, which improves the systematics. The single top production could also be studied from the CPT violation point of view.
In any case, a combination of different measurements will be necessary in order to consider the CPT violation as a serious explanation of any  unexpected data. 

Precision measurements in the top sector are just beginning and we do not know what surprises await us there. Here we have studied the exciting possibility of the observation of CPT violation. We have found a model independent parameter $R_{CPT}$ which parametrizes CPT violation, and shown that present constraints from Tevatron set $R_{CPT}< 10\%$. Future colliders will reduce this bound by one order of magnitude. From this study, the most promising channel is the lepton plus jets channel for top anti-top production. 

\begin{acknowledgments}
JARC acknowledges the hospitality and collaboration of workshop organizers
and conveners, and economical support from the NSF and Fulbright OLP. 
The work of JARC is supported in part by NSF grant No.~PHY--0239817, 
the Fulbright-MEC program, and the FPA 2005-02327 project (DGICYT, Spain). 
The work of AR is
supported in part by NSF Grant No.~PHY--0354993.  The work of FT 
is supported by the NSF grant No.~PHY-0355005.
\end{acknowledgments}


\begin{thebibliography}{99}

\bibitem{PDG}
S. Eidelman {\it et al.}, Phys. Lett. B {\bf 592} (2004) 1

\bibitem{P}
C.S. Wu {\it et al.}, Phys. Rev. {\bf 105} (1957) 1413 
R.L. Garwin {\it et al.} Phys. Rev. {\bf 105} (1957) 1415 

\bibitem{CP}
J.H. Christenson {\it et al.}, Phys. Rev. Lett. {\bf 13} (1964) 138 

\bibitem{T}
CPLEAR Coll., Phys. Lett. B {\bf 444} (1998) 43 

\bibitem{local} R. Jost, Helv. Phys. Acta {\bf 30} (1951) 39;
G.V. Efimov, Ann. Phys. {\bf 71} (1972) 466 

\bibitem{CPTmodels}
  D.~Colladay and V.~A.~Kostelecky,
  Phys.\ Rev.\ D {\bf 55} (1997) 6760;
  J.~P.~Hsu and M.~Hongoh,
  Phys.\ Rev.\ D {\bf 6} (1972) 256;
  C.~Ragiadakos and C.~Zenses,
  Phys.\ Lett.\ B {\bf 76} (1978) 61;
G.~Barenboim and J.~Lykken,
Phys.\ Lett.\ B {\bf 554} (2003) 73;
  Y.~A.~Sitenko and K.~Y.~Rulik,
  Eur.\ Phys.\ J.\ C {\bf 28} (2003) 405;
  R.~S.~Raghavan,
  JCAP {\bf 0308} (2003) 002;
  H.~Belich, J.~L.~Boldo, L.~P.~Colatto, J.~A.~Helayel-Neto and A.~L.~M.~Nogueira,
  Phys.\ Rev.\ D {\bf 68} (2003) 065030;
  C.Q.~Geng and L.~Geng,
  hep-ph/0504215

\bibitem{CDFdilept1} F.~Abe {\it et al.}, CDF Collaboration, 
Phys. Rev. Lett. {\bf 80} (1998) 2779 

\bibitem{CRT} J. A. R. Cembranos, A. Rajaraman and F. Takayama,
in preparation.

\bibitem{mtdetLHC} M.~Beneke et al., 
                   ``Top Quark Physics'', 
                   CERN-TH-2000-100, hep-ph/0003033, 
                   in: Standard Model Physics (and more) at the LHC, 
                   eds.\ G.~Altarelli and M.~Mangano, 
                   CERN, Geneva, 1999 [CERN-2000-004]. 

\bibitem{CDFl+jets} F.~Abe {\it et al.}, CDF Collaboration, 
Phys. Rev. Lett. {\bf 80} (1998) 2767

\bibitem{D0l+jets} B.~Abbott {\it  et al.}, D\O \ Collaboration,
Phys. Rev. D {\bf 58} (1998) 052001;
Phys. Rev. Lett. {\bf 79} (1997) 1197

\bibitem{Biernacik:2003xv}
  A.~Biernacik, K.~Kolodziej, A.~Lorca and T.~Riemann,
  Acta Phys.\ Polon.\ B {\bf 34} (2003) 5487

\end{thebibliography}
\end{document}